\documentclass[twocolumn,superscriptaddress,aps,amsmath,amssymb,floatfix,showpacs]{revtex4}
\usepackage{graphicx}
\usepackage{bm}
\begin{document}
\title{Multiple pre-edge structures in Cu K-edge of high T$_c$ cuprates \\
revealed by high resolution x-ray absorption spectroscopy}

\author{C. Gougoussis}
\affiliation{CNRS and Institut de Min\'eralogie et de Physique des Milieux condens\'es, 
case 115, 4 place Jussieu, 75252, Paris cedex 05, France}
\author{J.-P.~Rueff} 
\affiliation{Synchrotron SOLEIL, L'Orme des Merisiers, Saint-Aubin, BP~48, 91192 Gif-sur-Yvette Cedex, France} 
\affiliation{Laboratoire de Chimie Physique--Mati\`ere et Rayonnement, CNRS-UMR~7614, Universit\'e Pierre et Marie Curie, F-75005 Paris, France} 
\author{M. Calandra}
\affiliation{CNRS and Institut de Min\'eralogie et de Physique des Milieux condens\'es, 
case 115, 4 place Jussieu, 75252, Paris cedex 05, France}
\author{M. d'Astuto}
\affiliation{CNRS and Institut de Min\'eralogie et de Physique des Milieux condens\'es, 
case 115, 4 place Jussieu, 75252, Paris cedex 05, France}
\author{I. Jarrige}
\affiliation{Synchrotron Radiation Research Unit, Japan Atomic Energy Agency, 1-1-1 Kouto, Sayo, Hyogo 679-5148, Japan}
\author{H. Ishii}
\affiliation{National Synchrotron Radiation Research Center, Hsinchu 30076, Taiwan}
\author{A. Shukla}
\affiliation{CNRS and Institut de Min\'eralogie et de Physique des Milieux condens\'es, 
case 115, 4 place Jussieu, 75252, Paris cedex 05, France}
\author{I. Yamada}
\affiliation{Current Address: Department of Chemistry, Graduate School of  
 Science and Engineering, Ehime University, 2-5 Bunkyo-Cho, Matsuyama,  
 Ehime 790-8577, Japan}
\author{M. Azuma}
\affiliation{Institute for Chemical Research, Kyoto University, Uji, Kyoto  
 611-0011, Japan}
\author{M. Takano}
\affiliation{Current Address: Department of Chemistry, Graduate School of  
 Science and Engineering, Ehime University, 2-5 Bunkyo-Cho, Matsuyama,  
 Ehime 790-8577, Japan}

\date{\today}

\begin{abstract}
Using high resolution x-ray absorption spectroscopy and state-of-the-art electronic structure
calculations we demonstrate that the pre-edge region at the Cu K-edge of high T$_c$ cuprates is composed of several excitations
invisible in standard X-ray absorption spectra. We consider in detail the case
of Ca$_{2-x}$CuO$_2$Cl$_2$ and show that the many pre-edge excitations (two for c-axis polarization,
four for in-plane polarization and out-of-plane incident X-ray momentum) are 
dominated by off-site transitions and intersite hybridization. 
This demonstrates the relevance of approaches beyond the single-site model for 
the description of the pre-edges of correlated materials. 
Finally, we show the occurrence of a doubling of the main edge peak that is most visible when the polarization is
along the c-axis. This doubling, that has not been seen in any previous absorption data
in cuprates, is not reproduced by first principles calculations. We suggest that this peak
is due to many-body charge-transfer excitations, while all the other visible far-edge structures are
single particle in origin. Our work indicates that previous interpretations of the Cu K-edge X-ray absorption 
spectra in high T$_c$ cuprates can be profitably reconsidered.
\end{abstract}
\pacs{ 74.70.Ad, 74.25.Kc,  74.25.Jb, 71.15.Mb}

\maketitle

\section{Introduction}

The interpretation of X-ray absorption spectroscopy (XAS) at the Cu K-edge in high-T$_c$ cuprates, and especially of the pre-edge features, remains a challenging issue due to 
the strongly correlated nature of these materials. 
Understanding the nature of the pre-edge features is relevant since these excitations are
low in energy and probe the local environment of the absorbing atom. In several recent works 
\cite{shukla:077006,gougoussNiO,Kotani_Shukla08,Vankoxxx08} the presence of off-site excitations in 
transition metal compounds like NiO, cuprates and cobaltates was demonstrated. 
Off-site excitations can occur by a direct transition of a 1s core electron to empty electronic states of
the nearest neighboring atoms \cite{gougoussNiO}. This mechanism has a low intensity ($\sim 1\%$ of the edge jump) 
since the p-states of the absorbing atoms are not involved. More frequently the off-site excitations are mediated by the
on-site p-states of the absorbing atom that are very extended and therefore hybridize with neighboring sites.
When excitations due to intersite Cu (absorbing atom) 4p - Cu (other atom) 3d hybridization occur, the information is very
relevant since these features are typically weakly affected by core-hole attraction and thus they probe
the excitation in the absence of a core-hole in the final state. In some cases off-site transitions can be used to measure
the charge transfer gap, as  was recently suggested for NiO \cite{gougoussNiO}.
Therefore, a full understanding of pre-edge features and of their possible non-local nature leads to the
knowledge of the hybridization mechanism between different orbitals in the low energy region (up to some eV from the Fermi level)	
and of the position of the upper Hubbard band in the absence of a core-hole in the final state \cite{gougoussNiO}. 

In the case of high T$_c$ cuprates, and of correlated materials in general, 
the occurrence of many-body charge transfer excitations in the 
edge region has been widely debated   
\cite{tolentino_PhysRevB.45.8091,Bair_PhysRevB.22.2767,Kosugi_PhysRevB.41.131}.
Recently, using first principles calculations, we suggested \cite{gougoussis_uspp}
that  edge and far-edge structures previously attributed to charge transfer effects
were actually single particle in origin.
Thus most of the 
multi-determinant effects seen in photoemission seem to be suppressed in 
absorption, probably by the presence of a final-state in the XAS cross-section.

\begin{figure}[h]
\includegraphics[width=0.4\columnwidth]{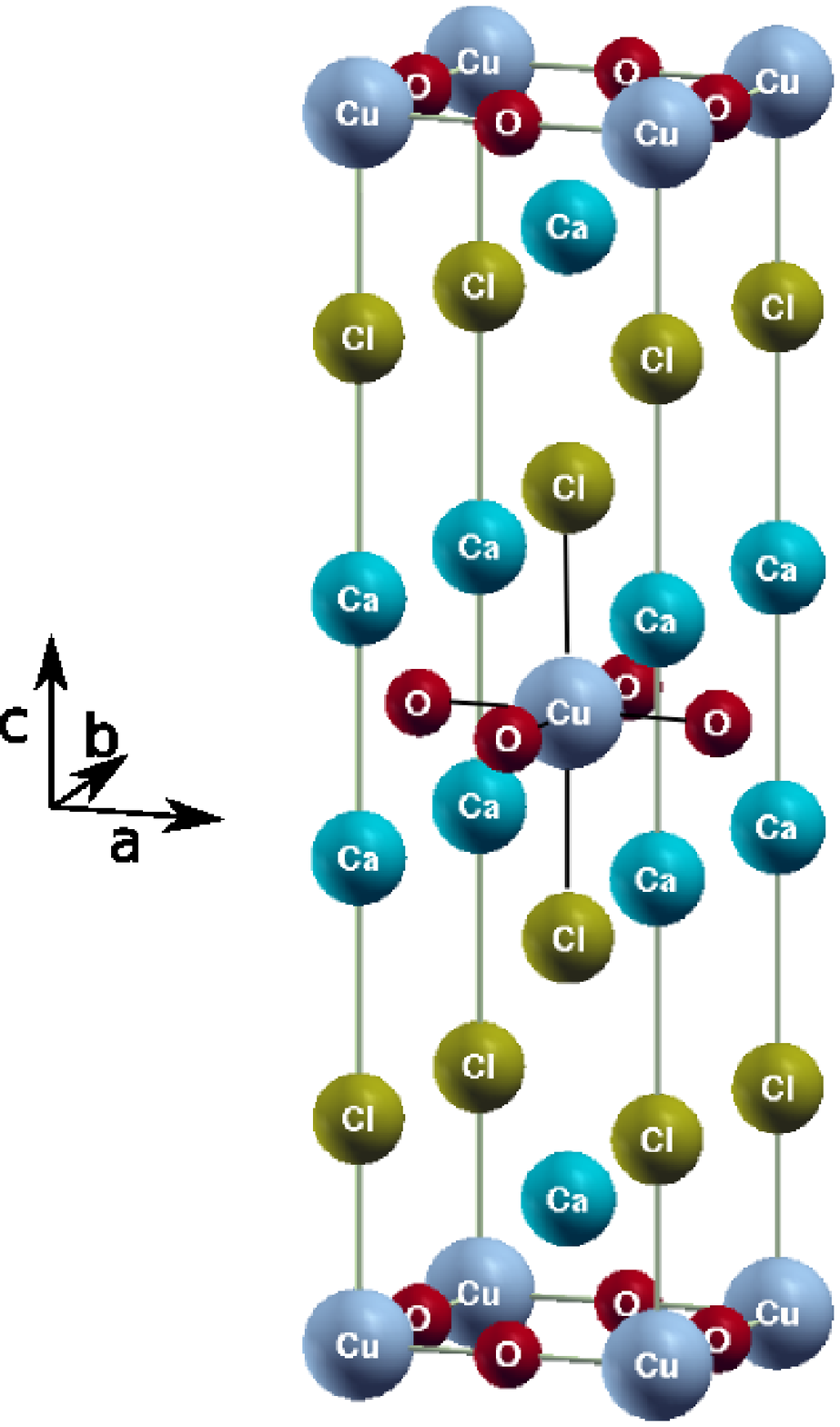}\includegraphics[width=0.4\columnwidth]{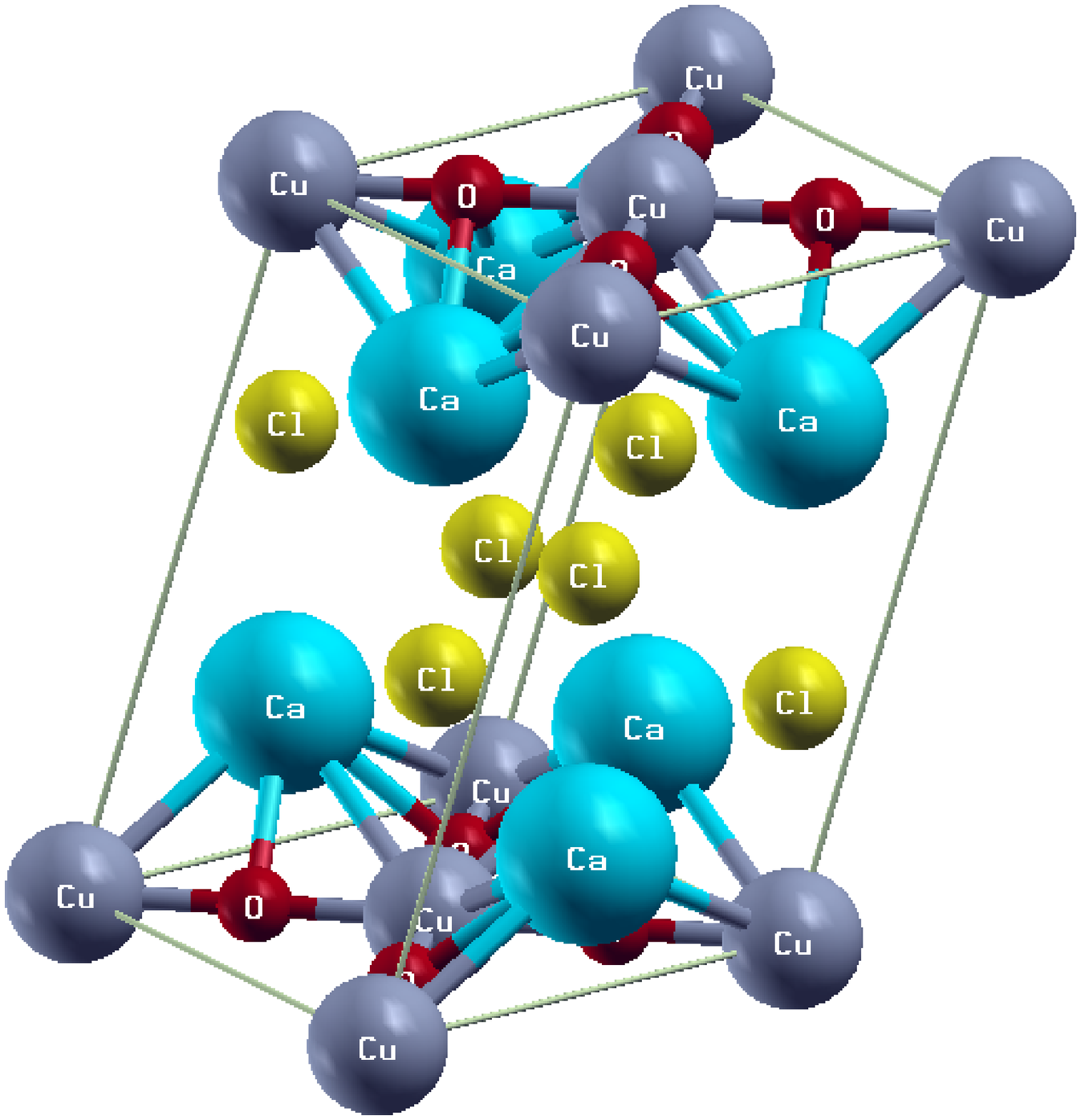}
\caption{(color online) Left: Crystal structure of Ca$_{2-x}$CuO$_2$Cl$_2$ 
\cite{xcrysden}. Right: 14 atoms monoclinic supercell used in the calculation.}
\label{fig:oxy0}
\end{figure}

In this work we present high resolution angular resolved partial-fluorescence yield (PFY) x-ray absorption 
measurements and first principle electronic structure calculations 
of the recently synthesized Ca$_{2-x}$CuO$_2$Cl$_2$ oxychlorides~\cite{yamada07}.
These materials form a new family of high Tc superconductors with a critical 
temperature peaking at 43 K at optimal nominal doping level (after post-annealing). 
The oxychlorides are appropriate for 
investigating the electronic properties of the CuO$_2$ plane.
They are, for example, well suited for calculations because 
of lesser structural disorder and the absence of rare earth atoms.
They also present peculiar features compared to the cuprates such as a 
checker-board pattern of charge inhomogeneities~\cite{hanaguri04} at odds 
with the stripes in the cuprates.
High resolution XAS allows us to identify new pre-edge  electronic excitations and a 
doubling of the main edge that are  not visible in previous XAS measurements
in high T$_c$ cuprates. Importantly we find that most of these excitations are 
single particle in nature. After identifying the 
single particle excitations and addressing the role of many-body charge-transfer  
effects in the XAS spectra, we come to the conclusion that the interpretation of Cu K-edge XAS in cuprates should be revisited.

Section \ref{sec:tech_det} presents the experimental and theoretical details,
and draws a comparison between the measured PFY-XAS spectra of Ca$_{1.8}$CuO$_2$Cl$_2$ and of La$_2$CuO$_4$.
In section \ref{sec:results}, we confront the experimental data with the calculated 
cross section and identify the single particle peaks using the projected density of 
states on selected electronic states in the presence of a core-hole in the 1s state. 
Finally we discuss the relevance of our results for the understanding of the 
empty state electronic structure in high T$_c$ cuprates.
\section{Technical details}
\label{sec:tech_det}

\subsection{Experiment}
The experiment was carried out at SPring-8 on the inelastic x-ray scattering (IXS) beamline BL12XU. The incident energy was selected by a Si(111) double crystal coupled to a high-resolution four-bounce Si(400) monochromator providing an estimated bandwidth of 220 meV. We have used well characterized superconducting single crystals of Ca$_{2-x}$CuO$_2$Cl$_2$ ($x=0.2$)\cite{yamada07}. To prevent degradation of the samples known to be highly hygroscopic, the crystals were covered with a grease suitable for low temperature measurements. The superconducting temperature of 35 K measured by SQUID confirms that the intrinsic quality was preserved. The sample orientation was verified by x-ray diffraction prior to measurements.

The absorption spectra were recorded in the partial fluorescence yield (PFY) mode using the beamline IXS spectrometer. The PFY-XAS consists of measuring the intensity variations of the Cu K$\alpha$ ($2p \rightarrow 1s$) emission line while scanning the incident energy through the Cu K-edge. With respect to conventional XAS, the PFY method yields an improved intrinsic resolution as the lifetime broadening effect in the PFY final state (with a $2p$ core hole) is considerably smaller than that of standard XAS at the K-edge (with a $1s$ core hole). The spectrometer was equipped with a 2 m bending radius Si(444) analyzer working at Bragg angle of 79.3$^\circ$ at the Cu K$\alpha$ energy (8048 eV) and a high flux silicon drift detector. The total energy resolution was estimated at ~250 meV from the width of the elastic line. The sample was mounted on a goniometer head with the $c$ axis (normal to the surface) in the horizontal plane. The PFY absorption spectra were acquired with the polarization of the incident photons set either parallel or perpendicular to the $c$ axis upon rotating the sample around a vertical axis. The spectra were corrected for self-absorption effect using the FLUO code developed by D. Haskel (Advanced Photon Source, Argonne).

\subsection{Calculation}
Ab initio calculations on Ca$_{2-x}$CuO$_2$Cl$_2$ are performed using spin polarized
density functional  theory (DFT) with 
PBE exchange-correlation \cite{PBE_PhysRevLett.77.3865}. 
We cross-check our results with the DFT+U approximation, with $U=9.6$ eV as
used in other cuprates \cite{gougoussis_uspp}. 
The calculation are performed with the 
XSPECTRA code \cite{xspectra,gougoussis_uspp} available under the GPL licence in the  
quantum-espresso package \cite{QE}. 
We use Ultrasoft pseudopotentials \cite{USPP_PhysRevB.41.7892}  for all 
atomic species. The 1s core-hole is  included in the pseudopotential of the absorbing 
atom. For the Cu pseudopotential we use non-linear core corrections 
\cite{NLCC} and do not include semicore states. 
It has been shown \cite{gougoussis_uspp} that the inclusion of the copper 
semicore states in the valence states is not necessary for an accurate description 
of XAS in cuprates.

For the XAS calculation we use a monoclinic supercell containing one CuO plane and 
14 atoms with experimental lattice parameters \cite{yamada07} (see Fig. \ref{fig:oxy0} ). 

 The magnetic state
of  Ca$_{2}$CuO$_2$Cl$_2$ has not been yet resolved, but for simplicity we assume
a collinear antiferromagnetism, similar to other cuprates.

The wavefunctions are expanded on a plane wave basis with a 30 Ry cutoff for 
the wavefunctions, and 400 Ry for the charge density. 
A uniform Monkhorst-Pack 6$\times$6$\times$2 k-points grid (with respect to the 
antiferromagnetric unit cell) is used for the charge density 
calculation, with a Methfessel-Paxton smearing of 0.05 Ry. 
The hole doping is treated by adding a compensating charge background, 
leading at $x=0.2$ to a nonmagnetic state in agreement with experiments.

The XAS cross section is calculated in the dipole approximation using 
an ultrasoft-based continued fraction approach   
\cite{gougoussis_uspp} and PAW reconstruction 
\cite{PAW_PhysRevB.50.17953}.
Two paw projectors are used in the absorbing atom pseudopotential. 
A uniform shifted 6$\times$6$\times$6 k-point grid is used for the continued 
fraction calculation. A 0.3 eV Lorentzian convolution, gradually increasing to 
1.5 eV after the edge is used to match the experimental width. 

\section{Results}
\label{sec:results}

\begin{figure}[t]
\centerline{\includegraphics[width=\columnwidth]{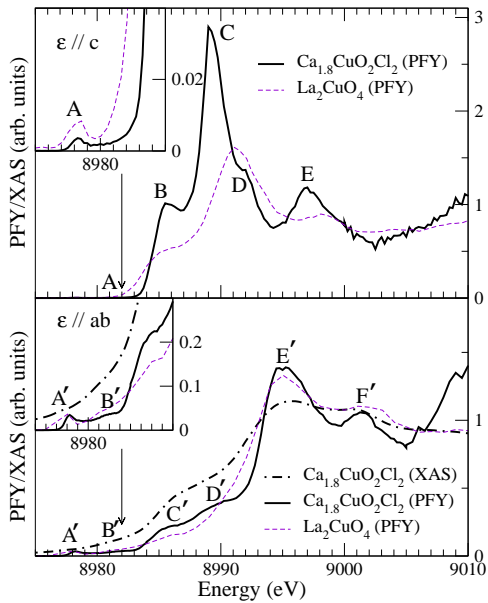}}
\caption{Comparison between the PFY-XAS Cu K-edge spectra of Ca$_{2-x}$CuO$_2$Cl$_2$ (solid lines) and La$_2$CuO$_4$ (dashed lines) for polarizations ${\bf \epsilon}\parallel c$ and ${\bf \epsilon}\parallel ab$ respectively in the top and bottom panels. In the case of Ca$_{2-x}$CuO$_2$Cl$_2$ the scattering geometries considered
are: (i) ${\bf k}$ parallel to the ab plane and ${\bf \epsilon}$ parallel to the c-axis,
labeled shortly ${\bf \epsilon}\parallel c$ and (ii)
${\bf k}$  parallel to the c-axis and
${\bf \epsilon}$ parallel to the ab plane,
labeled ${\bf \epsilon}\parallel ab$. In the case of La$_2$CuO$_4$ the geometries are
slightly different, as reported in ref. \cite{shukla:077006}. }
\label{fig:oxy4}
\end{figure}

\subsection{Experiment}

In figure \ref{fig:oxy4} we show the measured PFY-XAS data at the Cu K-edge in 
Ca$_{2-x}$CuO$_2$Cl$_2$ compared to previous PFY-XAS measurements \cite{shukla:077006} on
La$_2$CuO$_4$. In the case of Ca$_{2-x}$CuO$_2$Cl$_2$ the scattering geometries considered
are: (i) ${\bf k}$ parallel to the ab plane and ${\bf \epsilon}$ parallel to the c-axis,
labeled shortly ${\bf \epsilon}\parallel c$ and (ii) 
${\bf k}$  parallel to the c-axis and 
${\bf \epsilon}$ parallel to the ab plane,
labeled ${\bf \epsilon}\parallel ab$. In the case of La$_2$CuO$_4$ the geometries are
slightly different, as reported in ref. \cite{shukla:077006}.
For the ${\bf \epsilon}\parallel ab$ geometry we also plot 
standard XAS data in Fig. \ref{fig:oxy4}.

In the absence of a core-hole all d-states are filled except the 3d$_{x^2-y^2}$ 
orbital which is half-empty. 
However, the 3d$_{x^2-y^2}$ being planar, the quadrupolar part
is forbidden in both geometries. This is due to the fact
that the quadrupolar matrix element of the 
absorption cross section includes a term 
$({\bf k}\cdot {\bf r})({\bf \epsilon}\cdot {\bf r})$. In 
the ${\bf \epsilon}\parallel c$ geometry ${\bf \epsilon}\cdot {\bf r}=0$,
while in the ${\bf \epsilon}\parallel ab$ geometry 
${\bf k}\cdot {\bf r}=0$. So in both experimental
settings the matrix element is zero.

We first focus on the pre-edge region. At 8978 eV in ${\bf \epsilon}\parallel c$ geometry
an extremely small feature (labeled A in the following) is present, its intensity is 0.4 \% of the edge jump. 
A similar feature is present at the same
energy in the ${\bf \epsilon}\parallel ab$ geometry (labeled A'), however the intensity in this
direction is substantially larger, of the order of 3\% of the edge jump. Interestingly this
feature in the ${\bf \epsilon}\parallel ab$ geometry merges with a second broader band 
centered at 8982 eV (labeled B'). 
Given the weakness of some of these excitations, particularly
along the c-axis, we reconsider our published PFY-XAS Cu-K-edge La$_2$CuO$_4$ data 
to see if similar features occur \cite{shukla:077006}.
This is indeed the case, however these small features (in both geometries) 
were not discussed in our previous work \cite{shukla:077006}. 
We have not found any XAS measurements
\cite{Tranquada1987,Tranquada1991,tolentino_PhysRevB.45.8091,Kosugi_PhysRevB.41.131}
on cuprates displaying the A and A' features. We also performed XAS measurements
on the most interesting $\epsilon\parallel ab$ direction and found that 
the standard XAS spectrum shows a very broad tail extending down 
to the pre-edge region. None of the pre-edge features that are clearly resolved 
by PFY are visible here, as in the case of La2CuO4.
Furthermore the  B' at 8922 eV in the $\epsilon \parallel ab$ geometry
is usually assumed to be completely quadrupolar.
This is due to the lesser sensitivity 
of standard XAS data
when compared with PFY measurements.

\begin{figure}[t]
\centerline{\rotatebox{-90}{\includegraphics[width=\columnwidth]{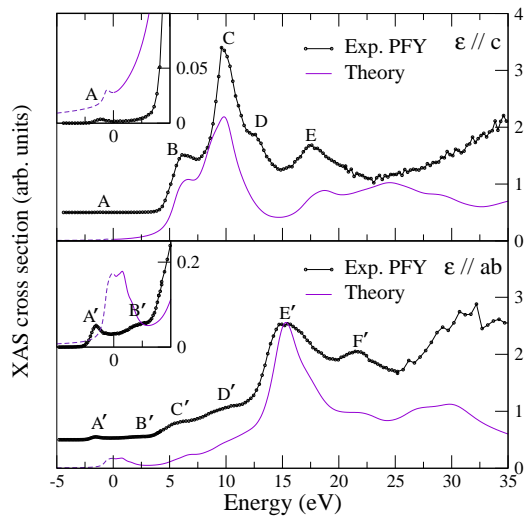}}}
\caption{Comparison between calculated Cu K-edge XAS (solid lines) and experimental (dots) PFY-XAS of Ca$_{2-x}$CuO$_2$Cl$_2$ ($x=0.2$) for both conditions of polarization ${\bf \epsilon}\parallel c$ and ${\bf \epsilon}\parallel ab$ respectively in the top and bottom panels. The core hole width of the calculation is 0.7 eV. The zero energy corresponds to 8979.2 eV}
\label{fig:oxy1}
\end{figure}

In  Ca$_{2-x}$CuO$_2$Cl$_2$  the first feature is probably
not more than 0.3 eV above the Fermi level and thus it is relevant for
its low energy physics. 
The other pre-edge features at higher energies (labeled B, B' and C' in fig. 
\ref{fig:oxy1}) are common to those of other cuprates. 

The structure in the absorption edge is similar in the two systems, apart from a 
remarkable difference. In the case of  Ca$_{2-x}$CuO$_2$Cl$_2$ with
  ${\bf \epsilon}\parallel c$ geometry the edge (peak of the whiteline) at 
8989 eV is accompanied by a
distinct shoulder at 8992 eV. 
In Ca$_{2-x}$CuO$_2$Cl$_2$ for ${\bf \epsilon}\parallel ab$ 
the edge seems substantially larger than the experimental resolution, 
suggesting the presence of a second peak in this case too.
In  La$_2$CuO$_4$ there is no evidence of such a splitting. However the poorer experimental resolution
we had in ref. \onlinecite{shukla:077006} could hinder
 the detection of a second peak. Consequently higher resolution
measurements are necessary to clarify if this splitting is a general feature of high
T$_c$ cuprates.

\subsection{Theoretical understanding}

\subsubsection{pre-edge region}

\begin{figure}[t]
\centerline{\rotatebox{-90}{\includegraphics[width=\columnwidth]{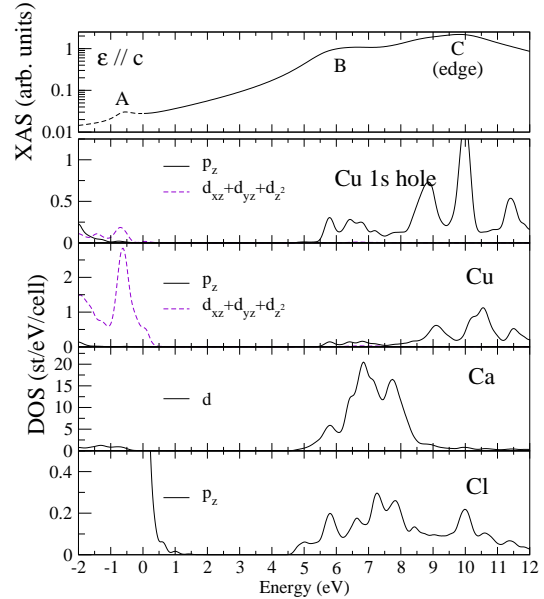}}}
\caption{Comparison between the calculated XAS cross section with $\boldsymbol{\epsilon}$ in the c direction and the density of states in the presence of a 
core-hole projected on selected orbitals. Note that
in this figure the XAS has been plotted in log scale to enhance the low intensity
low energy peaks.}
\label{fig:oxy2}
\end{figure}

\begin{figure}[t]
\centerline{\rotatebox{-90}{\includegraphics[width=\columnwidth]{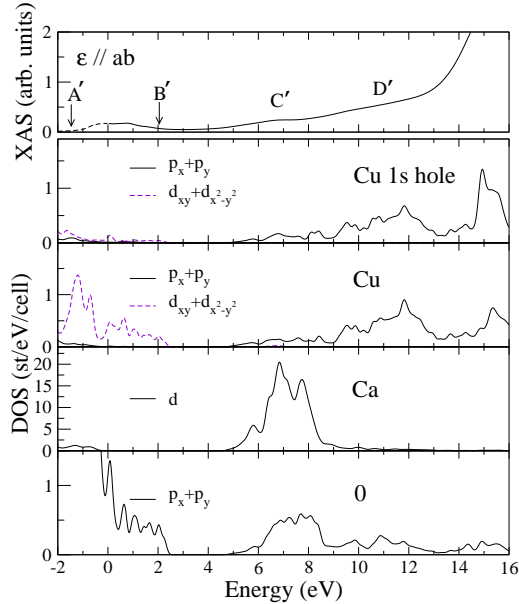}}}
\caption{Comparison between the calculated XAS cross section with $\boldsymbol{\epsilon}$ in the CuO plane direction and the density of states in the presence of a core-hole
projected on selected orbitals.}
\label{fig:oxy3}
\end{figure}

To understand the origin of the features seen in the absorption spectra, 
we carried out extensive DFT calculations for 
Ca$_{2-x}$CuO$_2$Cl$_2$. In the presence of a core-hole the d$_{x^2-y^2}$ state
around the absorbing atom is the only d state partially occupied.
In the pre-edge region  the situation is complicated by
the fact that the system is metallic and it is hard to isolate the
occupied states from the unoccupied ones, even as far as the dipolar part is concerned. 
As a consequence we have decided to shift the Fermi level 0.7 eV below that 
found in the supercell calculation (this is illustrated by the dashed line in
Figs.~\ref{fig:oxy1},\ref{fig:oxy2},\ref{fig:oxy3}).
This shift improves the agreement with experiments in the low-energy region.
However the drawback is that a small tail of the $d_{xz}+d_{yz}+d_{z^2}$ projected
density of states becomes unoccupied generating a quadrupolar term
(see Fig.\ref{fig:oxy3}). By direct calculation
of the XAS quadrupolar cross section we have verified that this quadrupolar term
is completely negligible with respect to the dipolar part emerging from the Fermi level
downshift.
Despite these difficulties and the strongly correlated nature of the system which is
a real challenge for density functional theory, we find good overall agreement with
experiments.

In the pre-edge region and ${\bf \epsilon}\parallel c$ geometry, 
we reproduce the low intensity peak  A  
as it can be seen in fig. \ref{fig:oxy1}.
The analysis based on the density of states projected on selected
orbitals in the presence of a core-hole in the final state (fig. \ref{fig:oxy2})
suggests that the peak is essentially due to an off-site transition to Cl 2p$_z$ states
and practically negligible quadrupolar component.  
This is analogous to what was found in NiO \cite{gougoussNiO}, namely 
a very low energy dipolar feature due to direct off-site transition
to nearby atoms.
The situation is not as favorable in the ${\bf \epsilon} \parallel ab$ geometry where
the broad band B' and the small feature A' are replaced in our calculation by
a more intense broad band (figs. \ref{fig:oxy1} and \ref{fig:oxy3}). This disagreement is a reminder that
DFT has limitations when it comes to the description of the low energy physics of high T$_c$ compounds.

On the other hand, the peaks C$^{'}$ and D$^{'}$ are correctly
reproduced. The C$^{'}$ peak is due to intersite hybridization 
between Cu 4p , O 2p$_{xy}$ and Ca d states, while the D$^{'}$ peak
is due to transitions to on-site Cu 4p states. 
Similarly to the C$^{'}$ peak, the B peak in
${\bf \epsilon}\parallel c$ geometry is due to intersite Cu 4p, Ca d and Cl p$_z$
states.
Thus all the pre-edge peaks A B, and C' are dominated by off-site transitions.

\subsubsection{Edge-to-far-edge region}

This energy region has been studied in detail in the past mainly due to the work of 
Tolentino {\it et al.} \cite{tolentino_PhysRevB.45.8091} and 
Kosugi {\it et al.} \cite{Kosugi_PhysRevB.41.131} suggesting that multi-determinant excitations
could occur. In particular, in ref. \cite{tolentino_PhysRevB.45.8091} it was proposed
that the edge peaks C and E' are due to transition to $|3d^{10} \underline{L} \rangle$ 
states while their supposed sattelites, E and F' are due to $|3d^9 \underline{L} \rangle$.
The original proposal was on Cu K-edge of La$_2$CuO$_4$. 
However in ref. \onlinecite{gougoussis_uspp} we showed that
this system is not well described by DFT in this region due to the difficulty in treating
La f states .
Nevertheless we concluded in
ref. \cite{gougoussis_uspp} that apart from a shift of the E peak to high energy,
generated by an incorrect treatment of La f states in DFT,
C,E,E', F' were most likely due to single particle excitations.
The present compound is ideal to verify this conclusion since no rare-earths are present.
Indeed we find that peaks C,E,E', F' are correctly reproduced confirming their single 
particle nature.

In the ${\bf \epsilon}\parallel c$ direction a peak labeled D is found. 
This peak is absent from our single particle calculation. 
This suggests that multi-determinant charge transfer effects could indeed occur 
in cuprates.
In order to test our conjecture one should consider the convolution between
hard X-ray Cu 1s photoemission data and first principles calculations. 
This procedure should give semi-quantitative agreement with the Cu K-edge XAS 
\cite{Nozieres69,Malterre91, mahan_conv, gunnarson_prb_1995, Hammoud87}. 
The difficulty in the present case is that there are no photoemission data 
for this system so measurements and calculations 
on other correlated compounds are needed.

\section{Conclusion}

In this work we have carried out Cu K-edge PFY-XAS measurements and state-of-the-art
first principles calculations in high T$_c$ cuprates Ca$_{2-x}$CuO$_2$Cl$_2$.
We have shown that rich structure due to dipolar transitions occurs in the pre-edge region. 
By performing electronic structure calculations we have shown that pre-edge 
peaks are dominated by off-site transitions.
Furthermore we have shown the occurrence
of new excitations in the low energy region, at less than 0.5 eV from the Fermi level.
In this energy region DFT calculations only partially capture the relevant physics
of the system since they fail in describing the in-plane excitations. Our work
demonstrates the need of going beyond single site cluster models and, for what
concerns correlation effects, beyond density functional theory, in order to
explain Cu K-edge XAS of cuprates.

Previous work 
\cite{tolentino_PhysRevB.45.8091,Bair_PhysRevB.22.2767,Kosugi_PhysRevB.41.131} 
invokes a strong influence of many body effects to explain the features observed in 
XAS K-edge spectra of high Tc materials. 
Notably the invocation of a correlated ground state and the rearrangement of 
the energy states due to the core hole effect has lead earlier authors to conclude 
that some of the most prominent peaks in XAS spectra are not due to single particle 
excitations. This interpretation is also inspired from core-level photoemission data 
but here we show that the situation is not analogous and that much of the XAS K-edge 
spectrum can in fact be explained invoking single particle excitations. 
Further, we find that many body effects due to charge transfer do occur 
but in a different way from earlier interpretations.
 
The ramifications of this reinterpretation go beyond XAS spectroscopy. 
It will suffice to note that RIXS spectra use the resonances found in XAS to 
excite particular intermediate states. 

\section{Acknowledgements}

Calculations were performed at the IDRIS supercomputing center (project 081202).
MCB and CG acknowledge fruitful discussions with F. Mauri, Ch. Brouder and
Ph. Sainctavit.

\end{document}